\documentclass[longbibliography,nofootinbib,twocolumn,showpacs,amsmath,amstex,amssymb,mathfonts,superscriptaddress,prl]{revtex4-2}

\usepackage{babel,calc,amsmath,amsthm,amssymb,graphicx,subfigure,xcolor,comment}
\usepackage{mathdots}
\usepackage[T1]{fontenc}
\usepackage[unicode=true]{hyperref}
\usepackage{braket}
\usepackage{dsfont}
\usepackage{booktabs}
\usepackage{mathtools}
\usepackage{stmaryrd}
\usepackage{natbib}

\usepackage[normalem]{ulem}
\hypersetup{
     colorlinks=true,       		
     linkcolor=orange,          	
     citecolor=blue,             
     urlcolor=cyan,          
 }

\newtheorem*{theorem*}{Theorem}

\newtheorem*{corollary*}{Corollary}

\newtheorem*{lemma*}{Lemma}

\newtheorem*{proposition*}{Proposition}
\theoremstyle{definition}

\newtheorem*{definition*}{Definition}
\theoremstyle{remark}

\newtheorem*{remark*}{Remark}

\newcommand\beq{\begin{equation}}
\newcommand\eeq{\end{equation}}
\newcommand\bal{\begin{aligned}}
\newcommand\eal{\end{aligned}}

\newif\ifdebug

\debugtrue

\ifdebug
\definecolor{zhliu}{rgb}{0.48, 0, 0.12}

\newcommand{\note}[1]{\textcolor{zhliu}{#1}}
\newcommand\delete{\bgroup\markoverwith{\textcolor{zhliu}{\rule[0.5ex]{2pt}{0.8pt}}}\ULon}

\else

\newcommand{\note}[1]{\ignorespaces}
\newcommand{\delete}[1]{\ignorespaces}

\fi

\begin{document}

\title{Emerging $(2+1)$D massive graviton in graphene-like systems}

\author{Patricio Salgado-Rebolledo}
\affiliation{Universit\'e Libre de Bruxelles and International Solvay Institutes, ULB-Campus Plaine CP231, B-1050 Brussels, Belgium}
\affiliation{Institute of Theoretical Physics, Wroc\l{}aw University of Science and Technology, 50-370 Wroc\l{}aw, Poland}

\author{Jiannis~K.~Pachos}
\affiliation{School of Physics and Astronomy, University of Leeds, Leeds LS2 9JT, UK}

\date{\today}

\begin{abstract}

Unlike the fundamental forces of the Standard Model the quantum effects of gravity are still experimentally inaccessible. Rather surprisingly quantum aspects of gravity, such as massive gravitons, can emerge in experiments with fractional quantum Hall liquids. These liquids are analytically intractable and thus offer limited insight into the mechanism that gives rise to quantum gravity effects. To thoroughly understand this mechanism we employ a graphene-like system and we modify it appropriately in order to realise a simple $(2+1)$-dimensional massive gravity model. More concretely, we employ $(2+1)$-dimensional Dirac fermions, emerging in the continuous limit of a fermionic honeycomb lattice, coupled to massive gravitons, simulated by bosonic modes positioned at the links of the lattice. The quantum character of gravity can be determined directly by measuring the correlations on the bosonic atoms or by the interactions they effectively induce on the fermions. The similarity of our approach to current optical lattice configurations suggests that quantum signatures of gravity can be simulated in the laboratory in the near future, thus providing a platform to address question on the unification theories, cosmology or the physics of black holes.

\end{abstract}


\maketitle

\section{Introduction}

An important open question in physics is that of observing quantum aspects of gravity. The coupling of gravity with matter is so weak that large, macroscopic masses are needed in order to generate an effect. Nevertheless, quantum effects are dominant in microscopic scales where gravity is negligible, thus, making quantum effects of gravity to be well beyond the reach of our current technology. This lack of any experimental evidence impedes our understanding of gravity at a fundamental level. For example, several often conflicting proposals exist for the quantisation of spacetime. Meanwhile, the $(3 + 1)$-dimensional graviton, the quantum particle that mediates gravity, is unrenormalisable, suffering from infinities in the theoretical level that cannot be removed. A possible way forward is to turn to condensed matter systems where gravitational effects could emerge at an effective level. As the couplings of such systems can be arbitrarily tuned then it would be possible to amplify the effect of geometric fluctuations in the simulated system and provide measurable signatures in proposed experiments.

The last decade has seen an increased interest in the geometric interpretation of condensed matter systems that emulate classical or quantum gravity effects. An interesting example is the emergence of massive gravitons in the $^3$He-B superfluid \cite{Volovik:2021wut}, where gravity emerges from fermionic bilinears after symmetry breaking, in analogy with the massless emergent gravitational field proposed by Diakonov to construct a lattice-regularized quantum gravity theory \cite{Diakonov:2011im}. Furthermore, of prominent interest is the exciting discovery that fractional quantum Hall liquids have collective excitations that are an analogue of gravitons. In particular, it has been demonstrated that the long wavelength limit of the Girvin-MacDonald-Platzman mode of the fractional quantum Hall effect \cite{PhysRevLett.54.581,PhysRevB.33.2481}, also known as the \emph{magnetoroton}, is properly described as a massive graviton excitation \cite{PhysRevLett.117.216403,PhysRevX.7.041032,PhysRevLett.120.141601}, which has been experimentally observed \cite{PhysRevLett.70.3983,PhysRevLett.84.546}. Nevertheless, fractional quantum Hall liquids are strongly interacting and thus analytically intractable. This intractability significantly limits our understanding of the mechanism responsible for the emergence of quantum gravity.

An alternative route is to engineer a system that effectively simulates quantum signatures of gravity. Such quantum simulations enable the realisation of a wide range of coupling regimes, thus enhancing properties that might be otherwise experimentally inaccessible. As an example, recent experimental advances in simulating quantum gauge theories offer insights into mechanisms that are not yet fully understood, like quark confinement \cite{zhou2021thermalization}. Several platforms exist where classical gravity emerges. For example, it is possible to produce non-trivial extrinsic geometry by deforming the shape of graphene-like systems~\cite{Iorio:2017vtw,PhysRevD.101.036021} or to create tunnelling coupling inhomogeneities that generate intrinsic geometries \cite{boada2011dirac,PhysRevB.101.245116}. Unfortunately, simulations of quantum gravity have not been achieved so far. Roadblocks exist at the conceptual level, such as in realising quantum fluctuations of spacetime, as well as at the practical level, such as in engineering the complex self-interactions present in gravitational theories. 

To resolve these problems we employ a simple and modular graphene-like system that can be analytically shown to give rise to massive gravitons in its low energy limit. In particular, we emulate Dirac fermions coupled to massive gravitons described by the Fierz-Pauli theory in 2+1 dimensions, which is known to posses two propagating degrees of freedom of helicity $\pm2$. To identify the right architecture we first consider the realisation of the Dirac field. We employ a two-dimensional honeycomb lattice configuration that describes $(2+1)$-dimensional Dirac fermions in its low energy limit. An effective background metric can be encoded in the couplings of the fermion lattice by making them position dependent~\cite{PhysRevB.101.245116}. Unlike extrinsically encoded geometry, which is hard to fluctuate, the intrinsically encoded one can be fluctuated by controlling the couplings of the model through auxiliary quantum fields~\cite{PhysRevLett.105.190403}. In our system we employ bosonic modes positioned at the links of the lattice to induce fluctuations of the gravitational field. These modes are coupled to the fermionic ones and are subject to density-density self-interactions in a particular way that gives rise to a semiclassical expansion of $(2+1)$-dimensional massive gravitons coupled to Dirac fermions. 

In recent years, $2+1$ dimensions have attracted great attention since they provide tractable models for gravity. Unlike $(3+1)$-dimensional gravity, pure $(2+1)$-dimensional Einstein gravity has no local propagating degrees of freedom and can be reformulated as a Chern-Simons theory \cite{Achucarro:1986uwr,Witten:1988hc} with boundary degrees of freedom given in terms of two-dimensional conformal field theories \cite{Brown:1986nw,Coussaert:1995zp,Strominger:1997eq,Barnich:2006av}. A mass term in the Einstein-Hilbert action introduces local gravitational degrees of freedom that imprint their effects on the local properties of the Dirac fermions. In higher dimensions, adding a mass for the graviton has been considered as a possibility to resolve the cosmological constant problem, as well as in the construction of theories for dark matter \cite{Babichev:2016bxi}. In four dimensions, the first consistent theory of massive gravity free of Boulware-Deser ghosts was found by de Rham, Gabadaze and Tolley \cite{PhysRevLett.106.231101}. Subsequently, massive generalisations of the Einstein-Hilbert action in 2+1 dimensions have been explored such as Topologically Massive Gravity, New Massive Gravity and Zwei-Dreibein Gravity \cite{Deser:1982vy,Bergshoeff:2009hq,Bergshoeff:2013xma}.

Apart from identifying a lattice model that effectively gives rise to massive gravitons in $2+1$ dimensions, our proposed architecture allows for quantum simulations with optical lattices. Ultra-cold atoms in optical lattices have proven to be an ideal system for probing interacting high-energy theoretical models or condensed matter physics that are otherwise inaccessible. From realising many-body localisation \cite{schreiber2015observation,PhysRevLett.114.083002,PhysRevLett.102.055301,choi2016exploring} and non-ergodicity in strongly correlated systems \cite{WinterspergerNP162020,scherg2021observing} to simulating lattice gauge theories \cite{Schweizer:2019lwx,yang2020observation,milsted2020collisions,mil2020scalable,zhou2021thermalization,Buser:2020cvn} or probing topological phases \cite{PhysRevLett.107.235301,PhysRevLett.113.045303,AtalaNP92013,flaschner2016experimental}, optical lattice technology is an invaluable component in advances of modern physics \cite{gross2017quantum}. Such simulations keep the promise of realising lattice gauge theories that can be used to illustrate quark confinement, a fundamental phenomenon that still remains a mystery \cite{Zohar_2015}. In particular, hexagonal optical lattices with fermionic or bosonic atoms have already been realised in the laboratory
~\cite{AtalaNP92013,flaschner2016experimental}.  Moreover, a shift in the paradigm of quantum simulations with cold atoms has happened with the suitable Yukawa-like coupling between fermionic and bosonic atoms for realising fluctuating gauge fields coupled to Dirac fermions~\cite{PhysRevLett.105.190403} that lead to numerous advances towards the quantum simulation of gauge theories~\cite{Schweizer:2019lwx,yang2020observation,milsted2020collisions,mil2020scalable,Buser:2020cvn}. Here, we present a novel boson-fermion interaction that encodes fluctuating gravitational fields coupled to Dirac fermions. The final necessary ingredient is the realisation of bosonic self-interaction terms that give rise to dynamics of the gravitational field. Such terms can be realised by employing Feshbach resonances that are routinely employed in experiments to realise a wide range of interactions between bosonic atoms~\cite{PhysRevLett.81.5109,PhysRevLett.103.265302,PhysRevA.99.033612,PhysRevLett.125.195302}. The Yukawa-like coupling between fermions and bosons causes the fermions to interact \cite{PRXQuantum.2.010325}, an effect that can be experimentally witnessed in the fermionic correlators through the violation of Wick's theorem~\cite{matos2021emergence}. The components necessary for this simulation, such as 2D Dirac fermions, mixtures of bosonic and fermions optical lattices and for controlling atomic intra and inter-species interactions, are routinely implemented with current experiments. Hence, optical lattices offer a key tool towards realising quatum properties of gravity in the laboratory.

\section{Dirac fermions coupled to gravitational fluctuations}
We first present the $(2+1)$-dimensional gravitational theory that we want to simulate. We choose the simplest possible theory that has interesting dynamics. To begin with, we choose a torsionless theory where all the dynamics comes from curvature. Moreover, General Relativity in three spacetime dimensions is known for possessing only global or boundary degrees of freedom, while no local propagating modes exist in the bulk similar to the theory in four spacetime dimensions. An interesting generalisation of this theory at the linear level is obtained by adding the Fierz-Pauli mass term. By endowing gravitons with a mass, the theory acquires two propagating degrees of freedom. This is the model we simulate with our lattice simulator.

To obtain spacetime geometries that can be realised with optical lattices we consider the spacetime metric to be in Gaussian form. This form provides a separation of time from space, thus giving a time evolution of the gravitational system that looks similar to the time evolution of the optical lattice, as dictated by the Schr\"odinger equation. This condition corresponds to choosing the spacetime coordinates in such a way that the metric $g_{\mu\nu}$ looks like
\beq
\label{gaussian}
ds^2=g_{\mu\nu}dx^\mu dx^\nu= -dt^2+ \left( l^2 \delta_{ij} +8\pi G h_{ij}\right) dx^i dx^j\,,
\eeq
where $G$ is Newton's constant, $l$ is a constant and $\mu=(t,i)$ denotes spacetime indices with $i=(x,y)$. To simplify the required optical lattice architecture we consider a gravity model where $h_{ij}$ is diagonal. This considerably simplifies the constraint structure of the theory and avoids the need of introducing extra relations between the lattice variables in order to preserve the constraints during time evolution. The coupling of the gravitation field with the Dirac fermions is best described in terms of the dreibein field $e^A_\mu$, given by $g_{\mu\nu}=\eta_{AB}e^A_\mu e^B_\nu$, where $A=(0,a=1,2)$ denotes Lorentz indices and $\eta={\rm diag}(-,+,+)$ is the metric of the local tangent Minkowski space. Parallel transport is then defined by the spin connection $\omega^A_\mu$ satisfying $\partial_\mu e^A_\nu - \Gamma^\rho_{\mu\nu}e^A_\rho +\epsilon^{A}_{\;BC}\omega^B_\mu e^C_\nu=0$,
where $\Gamma^\rho_{\mu\nu}$ is the affine Christoffel connection. We consider gravitational fluctuations of a flat geometry, $\bar e^A_\mu$. In this case the fluctuating gravitational field translates into fluctuations of the dreibein, $\xi^A_\mu$, and fluctuations of the spin connection, $v^A _\mu$, of a flat background, i.e:
\beq
\label{expeandomega}
e^A_\mu = \bar e^A_\mu+ 8\pi G \xi^A_\mu  \,,\qquad \omega^A_\mu = 8\pi G  v^A _\mu .
\eeq
Spacetime geometries of the form given in \eqref{gaussian}, with $h_{ij}$ diagonal, are described by
\beq
\label{gaugefixede}
\bar e^A_\mu= \begin{pmatrix}
1 & 0 \\
0 &l\, \delta^a_i 
\end{pmatrix}
,\quad 
\xi^A_\mu= \begin{pmatrix}
0 & 0 \\
0 & \xi^a_i 
\end{pmatrix},\quad
\xi^a_i = \begin{pmatrix} 
 \xi^1_x & 0\\
 0 & \xi^2_y 
\end{pmatrix},
\eeq 
where $\xi^a_i$'s are the spatial dreibein fluctuations. 
The metric fluctuations are then given by $h_{ij}= l\left(\delta_{ai}  \xi^a_j +  \delta_{aj}\xi^a_i\right)$. Furthermore, we consider torsionless geometries, for which the spin connection perturbation $v_\mu^A$ can be expressed in terms of derivatives of $\xi^A_\mu$ (see Appendix \ref{AppA} for details).

In the following we consider a gravitational model described by the action 
\beq
\label{fullaction}
S[\psi,\xi]=S_\text{Dirac}[\psi,\xi]+S_\text{gr} [\xi],
\eeq
where $S_\text{Dirac} $ is the action for a Dirac spinor $\psi$ that includes the coupling of the geometry $\xi^a_{i}$ to the fermionic current, whereas $S_\text{gr} $ is a purely gravitational action that describes the spatial dreibein fluctuations $\xi^a_{i}$ in a flat background geometry. 

\subsection{Fermionic action}

The fermion action describes a massless Dirac field on curved space
\beq\label{gfgeometry}
S_\text{Dirac}=\frac{i }{2} \int  d^3 x |e| \left( \bar \psi e^\mu_A \gamma^A \overrightarrow D_\mu \psi- \bar  \psi \overleftarrow D_\mu e^\mu_ A \gamma^A \psi   \right),
\eeq
where $\bar\psi=\psi^\dagger \gamma^0$, and the covariant derivative acting on fermionic fields is defined as $\overrightarrow D_\mu = \overrightarrow\partial_\mu+ \omega_\mu$ and
$\overleftarrow D_\mu = \overleftarrow\partial_\mu  - \omega_\mu$ with $\omega_\mu =\frac{1}{4}\epsilon_{ABC} \omega^A_\mu \gamma^B\gamma^C$. In order to compare the fermionic action $S_\text{Dirac}$ with the one coming from the optical lattice simulation, we rescale the corresponding spinor $\psi$ as $\psi \longrightarrow  \psi/{\sqrt{| e|} }$, so that they both satisfy flat anti-commutation relations \cite{PhysRevB.98.064503}. Next we split the temporal and spatial indices and implement the semiclassical expansion \eqref{expeandomega}. For small Newton's constant the resulting fermionic action to linear order in $G$ is given by
\beq\label{factionfinal}
\bal
S_\text{Dirac}[&\psi,\xi] = i \int  d^3 x  
 \left[\bar \psi  \gamma^0 \dot \psi+\bar \psi  \gamma^i \partial_i \psi  \right]\\
 & -\frac{8i\pi G}{l^2} \int  d^3 x  \left[
  \xi^i_a \bar \psi   \gamma^a \partial_i \psi+\frac{1}{2}
 \partial_i \xi^i_a  \bar \psi   \gamma^a \psi \right],
\eal
\eeq
where we have defined the inverse field $\xi^i_a =\delta^i_b \delta^a_j \xi^b_j$.

\subsection{Gravitational action}

In order to describe the dynamics of the gravitational field we start with the Palatini action for $e^A$ and $\omega^A$ given by 
\beq\label{Pal}
S_\text{gr} ={1 \over 8\pi G}  \int d^3 x \epsilon^{\mu\nu\rho}\,e^A_\mu  \left(\partial_\nu \omega_{A\rho} + \frac{1}{2} \epsilon_{ABC}\; \omega^B_\nu \omega^C_\rho \right).
\eeq
As before, we consider the semiclassical expansion of $S_\text{gr}$ with zero torsion and background curvature. 
The dominant order in the $G$ expansion is then given by the massless Fierz-Pauli action for $h_{\mu\nu}=\bar e_{A\mu}\xi^A_{\nu}+\bar e_{A\nu}\xi^A_{\mu}$ (see Appendix \ref{AppB} for details). The particular action and geometries are suitably chosen so that the resulting Hamiltonian can be directly modelled with ultra-cold atoms in optical lattices. Moreover, considering a metric in Gaussian coordinates \eqref{gaussian} is convenient at the quantum level, since it allows to eliminate the so-called \emph{conformal divergences} in the graviton path integral \cite{Dasgupta:2001ue}. 

The action \eqref{Pal} describes a topological theory with no propagating degrees of freedom. In order to have local degrees of freedom, akin to $(3+1)$-dimensional General Relativity, we introduce a mass term for the gravitational field. Here, we consider massive gravitons $\xi^A_\mu$ described by the Fierz-Pauli theory \cite{fierz1939relativistic}, obtained by adding the mass term 
\beq\label{FPmass}
{4\pi G\mu^2}\epsilon^{\mu\nu\rho}\epsilon_{ABC}\, \bar e^A_\mu \xi^B_\nu  \xi^C_\rho \,,
\eeq
in the Lagrangian \eqref{Pal} after linearisation. Combining \eqref{expeandomega}, \eqref{Pal} and \eqref{FPmass} finally gives (see Appendix \ref{AppB} for details)
\beq\label{gravactionfinal}
\bal
S_\text{gr} [\xi]= -4\pi G \int&d^3 x \epsilon^{ij}\epsilon_{ab}  \bigg(   \dot \xi^a_{i} \dot \xi^b_{j} 
- \mu^2 \xi^a_i \xi^b_j\bigg),
\eal
\eeq
which is the effective massive gravitational action, with $\xi^a_{i}$ given by \eqref{gaugefixede}. Note that we have introduced the mass term \eqref{FPmass} by hand. It is known that the Fierz-Pauli action in 2+1 dimensions can be obtained from New Massive Gravity \cite{Bergshoeff:2009hq,Bergshoeff:2009aq}, a full gravitational theory with quadratic-in-curvature terms that possesses two local propagating degrees of freedom of helicity $\pm2$. Alternatively, a different gravitational theory, whose weak field limit leads to the same massive Fierz-Pauli action has been proposed in \cite{Visser:1997hd}.

\subsection{Total Hamiltonian}

In order to determine the optical lattice configuration required to simulate this gravity model, we need first to obtain its Hamiltonian. The Hamiltonian corresponding to the action \eqref{fullaction} is given by (for details, see Appendix \ref{AppC})
\beq
\label{GQFThamiltonian}
H= \int  d^2 x\; \left[ \psi^\dagger \; h({\boldsymbol p})\; \psi  + \mathcal H_\text{gr}\right],
\eeq
where the single particle Hamiltonian $h({\boldsymbol p})$ reads
\beq
\label{singpart}
h({\boldsymbol p})=\frac{\gamma^0}{l} \left( \delta^i_a\gamma^a -\frac{8\pi G}{l } \xi ^i_a \gamma^a \right)(-i \partial_i)  +\frac{4i\pi G}{l^2}
 \partial_i \xi^i_a  \gamma^0  \gamma^a ,
\eeq
and the gravitational Hamiltonian $\mathcal{H}_\text{gr}$ equals
\beq
\label{gravhamiltonian}
 \mathcal{H}_\text{gr} =-  \frac{1}{16\pi G}\epsilon_{ij} \epsilon^{ab} \pi_a^i  \pi_b^j+{4\pi G\mu^2}\epsilon^{ij} \epsilon_{ab}  \xi^a_i \xi^b_j .
\eeq
Here $\pi^i_a =\text{diag}(\pi^x_1, \pi^y_2)$ is the canonical momentum conjugate to $\xi^a_i$ given in \eqref{gaugefixede}. The geometric fluctuations and their conjugate momenta can be expressed as 
\beq
\bal
&\xi^1_x= \frac{1}{\sqrt 2}(q_1^\dagger +q_1)\,,\\&\xi^2_y= \frac{1}{\sqrt 2}(q_2^\dagger +q_2)\,,
\eal
\,\qquad
\bal
&\pi_1^x= -\frac{i}{\sqrt 2 }(q_1^\dagger -q_1)\,,\\&\pi_2^y=-\frac{i}{\sqrt 2 }(q_2^\dagger -q_2)\,,
\eal
\eeq
where the operators $q_a$ and $q^\dagger_a$ $(a=1,2)$ satisfy the bosonic commutation relations $[q_a(x), q_b^\dagger(y) ]=\delta_{ab}\delta^{(2)}(x-y)$ and $[ q_a(x), q_b(y) ]=0=[q_a^\dagger(x), q_b^\dagger(y) ]$. It is in terms of these quantum operators that we can establish a map between \eqref{GQFThamiltonian} and an optical lattice Hamiltonian.

\begin{figure}
\centering
\includegraphics[width=\columnwidth]{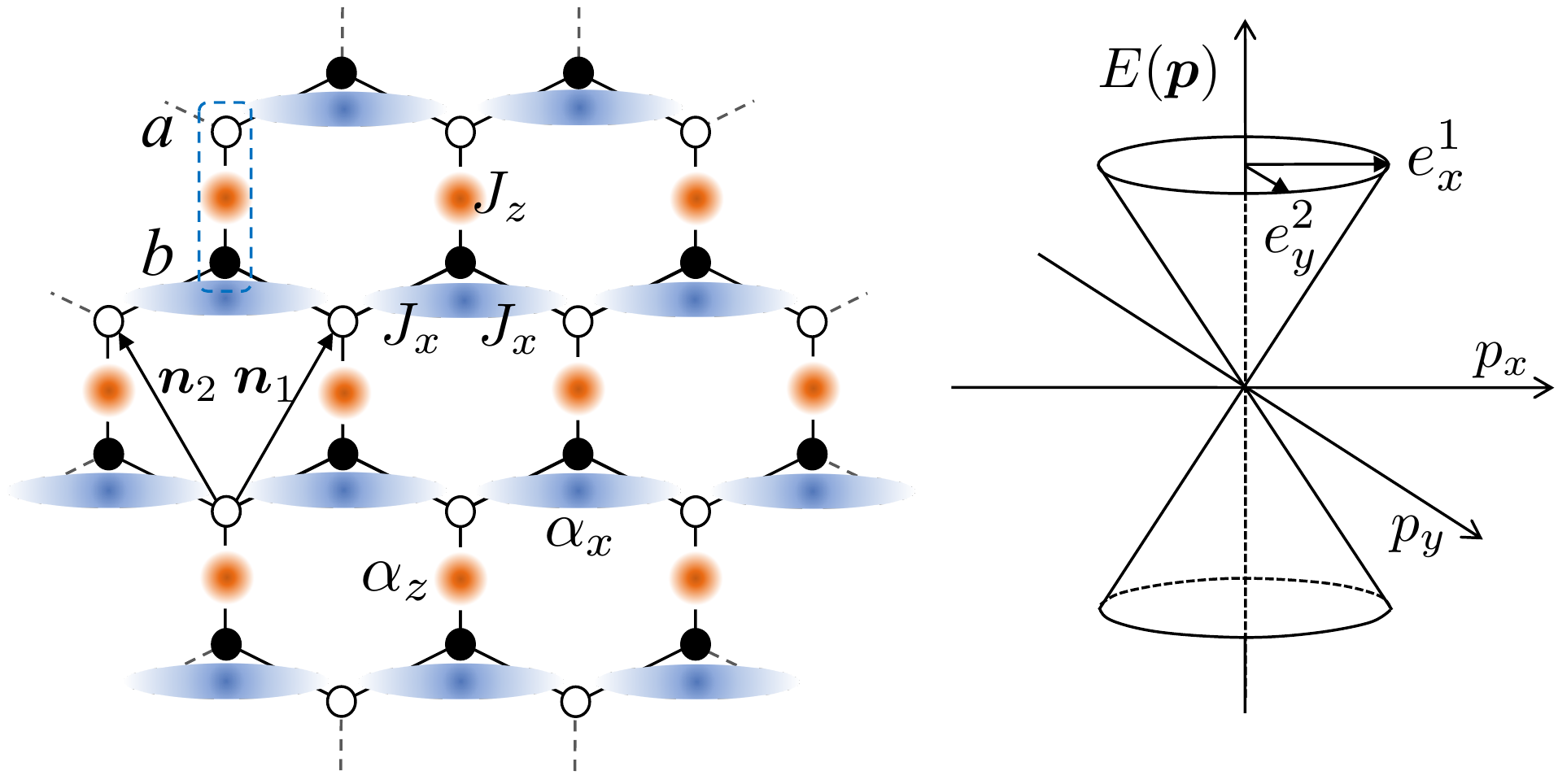}
\caption{(Left) The optical lattice configuration with fermionic and bosonic atoms. The unit cell of the honeycomb lattice (dashed box) comprises two fermionic modes $a$ and $b$. The fermionic atoms tunnel along the three different directions of the trivalence lattice with couplings $J_x$, $J_y(=J_x)$, and $J_z$. The vectors ${\boldsymbol n}_1 =(\sqrt{3}/2,3/2)$ and ${\boldsymbol n}_2 =(-\sqrt{3}/2,3/2)$ transport between different unit cells.
(Right)~In the low energy limit the dispersion relation $E({\boldsymbol p})$ of the fermions is given by Dirac cones.
Non-equal tunnelling couplings, $J_x$ and $J_z$, cause the Dirac cone to be deformed, with its geometry encoded in the dreibein components $e^1_x$ and $e^2_y$, effectively describing background gravity. Bosonic modes $\alpha_x$ and $\alpha_z$, describing Bose-Einstein condensates, are inserted that control the fermionic tunnelling couplings with their populations. Fluctuations of the condensates simulate fluctuations in $e^1_x$ and $e^2_y$, i.e. gravitational fluctuations.}
\label{fig:lattice}
\end{figure}

\section{Optical lattice simulator}

\begin{figure}
\centering
\includegraphics[width=0.9\columnwidth]{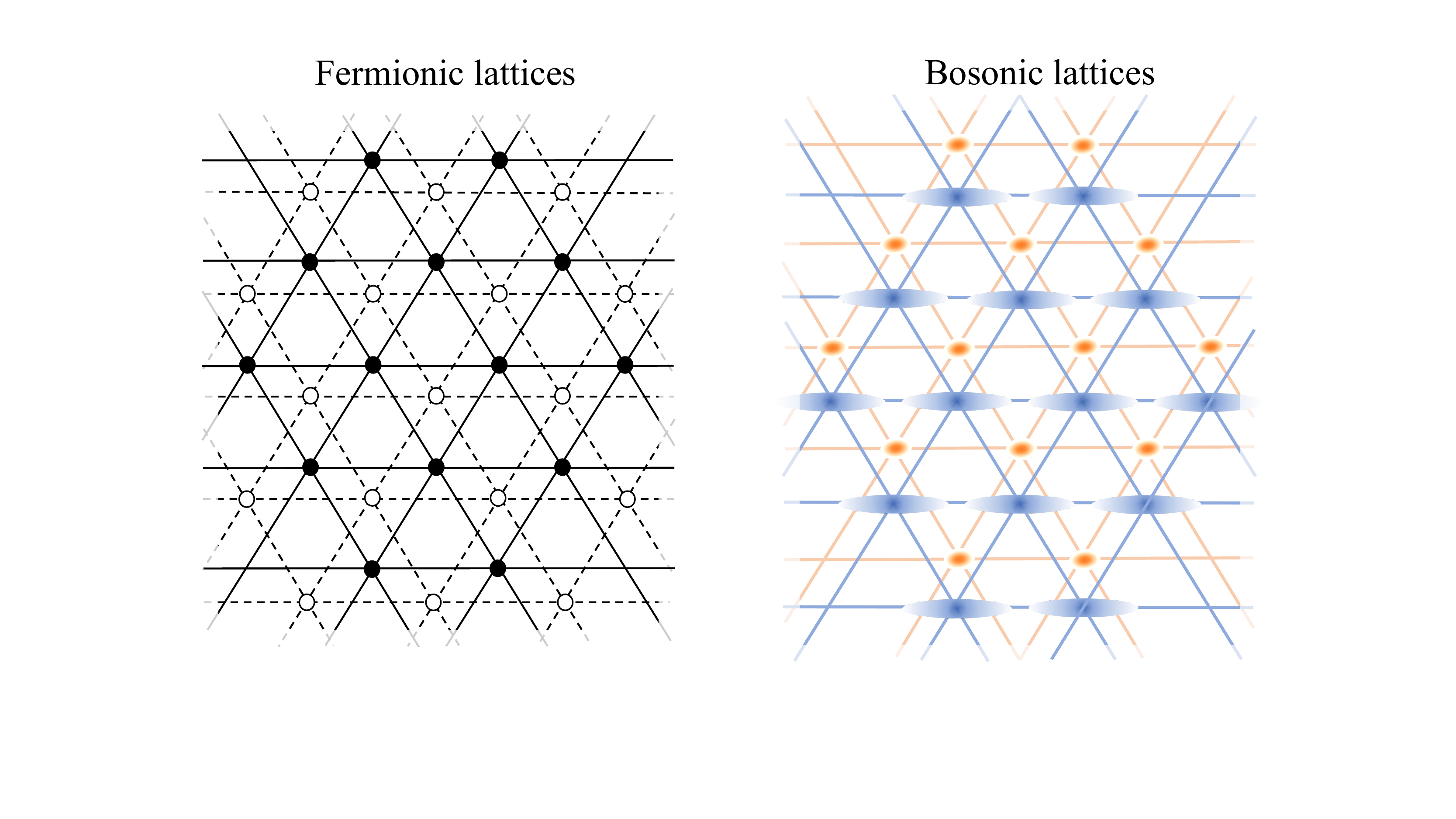}
\caption{The optical lattice configuration of Figure~\ref{fig:lattice} is obtained by superposing two pairs of triangular lattices, a fermionic pair (Left) and a bosonic pair (Right). The pair of lattices that hosts fermionic modes $a$ and $b$ gives rise to the honeycomb lattice model when nearest neighbour tunnelling couplings are activated. The pair of bosonic lattice modes, $\alpha_x$ and $\alpha_z$, is positioned at the links of the honeycomb lattice. The $x$-confinement of the $\alpha_x$ modes is made weaker to overlap with both $J_x$ and $J_y$ links of the honeycomb lattice (see Figure~\ref{fig:lattice}). By activating interactions between the bosonic and fermionic atoms, $\Delta_m \alpha^\dagger_m \alpha_m a^\dagger_{\boldsymbol i} b_{\boldsymbol k}$, the population of the bosonic modes at site $m$ controls the fermionic tunnelling between sites ${\boldsymbol i}$ and ${\boldsymbol k}$, thus giving rise to the effective interaction between Dirac fields and fluctuating geometry.}
\label{fig:superlattices}
\end{figure}

We now design an optical lattice configuration that simulates the quantum gravity model given by \eqref{GQFThamiltonian} in its low energy limit.
We first present a configuration that gives rise to Dirac fermions in a fixed background geometry encoded in the tunnelling couplings of the lattice \cite{PhysRevB.101.245116}. Then we introduce link quantum variables $d_m$ $(m=x,z)$ that fluctuate this geometry. Consider a two-dimensional optical lattice with honeycomb configuration where fermionic atoms, $a$ and $b$, live at its vertices, as shown in Figure~\ref{fig:lattice}(Left). The fermions are subject to the tunnelling Hamiltonian
\beq
\label{OLH}
H_\text{latt} = \sum_{\boldsymbol i} \big[J_x(\boldsymbol i) a_{\boldsymbol i}^\dagger b_{\boldsymbol i+\boldsymbol n_1} + J_y(\boldsymbol i)  a_{\boldsymbol i}^\dagger b_{\boldsymbol i+\boldsymbol n_2} +J_z(\boldsymbol i)  a_{\boldsymbol i}^\dagger b_{\boldsymbol i}\big] +\text{h.c.},
\eeq
where ${\boldsymbol i}=(i_x,i_y)$ gives the position of unit cells on the lattice, the couplings of the first two terms are equal ($J_x=J_y$), and both $J_x$ and $J_z$ are in general position dependent. At half-filling, the low energy sector of the model is described by the Dirac Hamiltonian $H_\text{latt} \approx \int  d^2 x\; 
 \psi^\dagger \;\tilde h({\boldsymbol p}) \;\psi$ \cite{PhysRevB.101.245116} with (see Appendix \ref{AppD} for details)
\beq
\bal
\tilde h({\boldsymbol p})&=  -\frac{\sqrt3i}{2} \sqrt{4J_x^2-J_z^2}\,\gamma^0\gamma^1\partial_x  
- \frac{3i}{2}J_z \gamma^0\gamma^2\partial_y\\
&+ \frac{\sqrt3i}{2} \partial_x \left(\sqrt{4J_x^2-J_z^2}\,\right)\gamma^0\gamma^1+ \frac{3i}{2}\partial_y J_z \gamma^0\gamma^2.
\eal
\label{eqn:Diracsim}
\eeq
The coefficients of $\gamma^0\gamma^1(-i\partial_x)$ and $\gamma^0\gamma^2 (-i\partial_y)$ play the role of the diagonal space components of the dreibein, $e^1_x$ and $e^2_y$, respectively, as shown in Figure~\ref{fig:lattice} (Right). To introduce quantum fluctuations in the $J$ couplings, and thus in the corresponding dreibeins, we insert bosonic modes $\alpha_x$ and $\alpha_z$ at the edges of the lattice, as shown in Figure~\ref{fig:lattice} (Left). To realise this configuration of atomic boson-fermion mixture we need two pairs of triangular lattices, one fermionic and one bosonic, as shown in Figure~\ref{fig:superlattices}. Triangular lattices can routinely be created in the laboratory for bosonic~\cite{Becker_2010} and fermionic~\cite{Yamamoto_2020} atoms. Optical lattices with interacting mixtures of bosonic and fermionic atoms have been realised and their rich physics has been extensively investigated~\cite{Truscott_2001,akatsuka2008optical,sugawa2011interaction}.

When the bosonic and fermionic lattices shown in Figure~\ref{fig:superlattices} are superposed, the bosonic modes $\alpha_x$ and $\alpha_z$ control the tunnelling of fermions from site $\boldsymbol i$ to site $\boldsymbol k$ in the $m(=x,z)$ direction, through the interaction $\Delta_m \alpha^\dagger_m \alpha_m a^\dagger_{\boldsymbol i} b_{\boldsymbol k}$~\cite{PhysRevLett.105.190403}. We take the $\alpha_j$ modes to correspond to a bosonic condensation with particle density $D_m$ and quantum fluctuations $d_m$, i.e.
\beq\label{defalpham}
\alpha_m(\boldsymbol i) = D_m(\boldsymbol i) + d_m(\boldsymbol i) ,
\eeq
where $[d_m(\boldsymbol i),d_m^\dagger(\boldsymbol j)]=\delta_{\boldsymbol i \boldsymbol j}$, for $m=x,z$. In the weak fluctuation limit $\langle d^\dagger_m d_m\rangle \ll D_m^2$, the interactions between bosons and fermions give rise to tunnelling couplings of the form $J_m= \Delta_m D_m(D_m +d_m^\dagger +d_m)$.
If we choose the optical lattice parameters as
\beq\label{DxDeltax}
D_{x}=\sqrt{2} D_z = -{l\over 4 \pi G}\,,\qquad \Delta_x={\Delta_z \over 2} = \frac{32\pi^2G^2}{3 l^3 },
\eeq
with the operator redefinition $q_1= {2\sqrt2} d_x/3 - d_z/3$ and $q_2=d_z$ that preserves bosonic commutation relations, we find that, to linear order in $G$, the optical lattice Hamiltonian \eqref{OLH} is mapped to the field theory one \eqref{singpart} (for details, see Appendices \ref{AppC} and \ref{AppD}). Constant tunnelling terms can be added in \eqref{OLH} to arbitrarily tune the values of the densities $D_m$.

The self-interaction terms \eqref{gravhamiltonian} of the gravitational Hamiltonian can be obtained from the following purely bosonic interactions
\beq
\bal
\label{eq:bos}
\mathcal H_\text{boson} =& \frac{ 1}{24\pi G}(\alpha_z^\dagger -\alpha_z)\left(\sqrt2 (\alpha_x^\dagger -\alpha_x)-\frac{1}{2}
(\alpha_z^\dagger -\alpha_z)\right)\\
&+\frac{8\pi G \mu^2}{3}\left(\alpha^\dagger_z \alpha_z+ 
\alpha^\dagger_x \alpha_x\right)\\
& 
-\frac{ 256\pi^3G^3\mu^2}{3 l^2} \alpha^\dagger_z \alpha_z \left(\alpha^\dagger_x \alpha_x 
-\frac{1}{2}\alpha^\dagger_z \alpha_z\right),
\eal
\eeq
restricted to  the weak fluctuation regime of $\alpha_m$.

\subsection{Realisation of interactions}

The quantum simulation of Dirac fermions coupled to fluctuating gravity requires the realisation of \eqref{GQFThamiltonian} with the fermionic part coupled to the bosons given by \eqref{singpart} and the self-interactions of bosons \eqref{eq:bos}. The first component is the realisation of the $(2+1)$-dimensional Dirac dispersion relation of the form \eqref{eqn:Diracsim}. This has already been achieved in the laboratory with ultracold fermionic atoms by several experimental groups \cite{tarruell2012creating,duca2015aharonov,li2016bloch}. Of much interest is the possibility the optical lattices offer in running the couplings at will thus allowing the effective encoding of background geometry \cite{PhysRevB.101.245116}. Introducing fluctuations in the background geometry can be achieved with optical lattices by coupling bosonic and fermionic species together, as shown in Figure \ref{fig:lattice}(Left). This method is similar to the minimal coupling realisation with optical lattices \cite{PhysRevLett.105.190403}, suitably adapted here to encode the coupling between Dirac fermions and dreibeins. It incorporates Bose-Einstein condensates with weak boson number fluctuations. When positioned at the links of the honeycomb lattice they couple with the tunnelling fermions producing the desired boson-fermion couplings dictated by \eqref{singpart}. Note that the tunnelling couplings of fermions, $J_x$, $J_y$ and $J_z$, and the condensation particle densities $D_m$ can be routinely tuned and controlled in an experiment by controlling the laser intensity of the optical lattices and the frequency of the magnetic or optical trapping of the condensate, respectively.

\begin{figure}
\centering
\includegraphics[width=0.8\columnwidth]{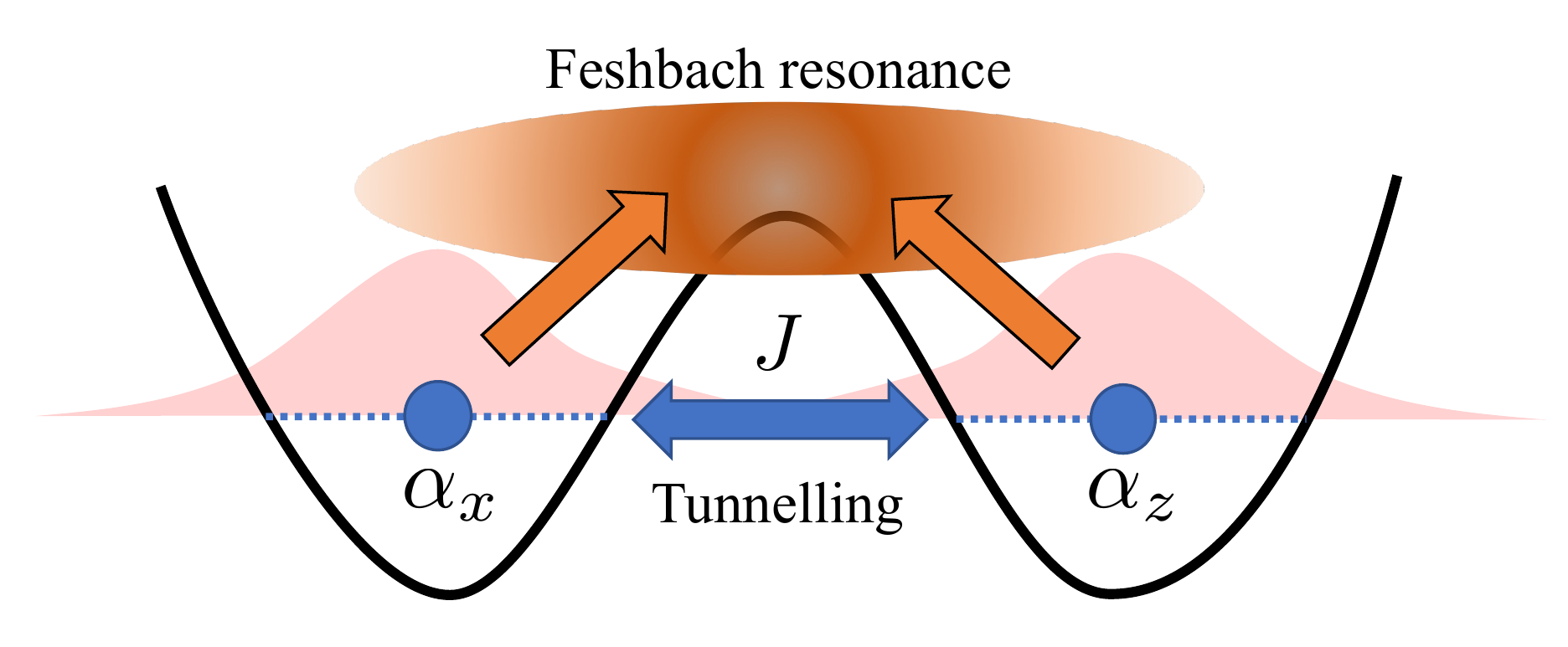}
\caption{Realisation of the self-interacting terms of the gravitational bosonic modes $\alpha_x$ and $\alpha_z$. Tunnelling couplings of the form $J(\alpha_x^\dagger \alpha_z +\alpha_z^\dagger \alpha_x)$ are produced by the overlap of the wavefunctions of the modes that sit in different wells. The tunnelling coupling $J$ is controlled by the barrier between the potential wells tuned by the laser intensity of the optical lattices. Feshbach resonances are produced by the elastic scattering collisions between $\alpha_x$ and $\alpha_z$ atoms with a rate that can be easily tuned by a magnetic field.}
\label{fig:bosons}
\end{figure}

The final ingredient is the realisation of the gravitational self-interacting terms presented in \eqref{eq:bos}. This Hamiltonian includes chemical potentials as well as tunnelling, pairing and interacting terms between the $\alpha_m$ bosonic atoms. Chemical potentials can be easily tuned with great accuracy by controlling the populations or the trapping potential of the cold atoms. Tunnelling terms between two different bosonic modes, such as $\alpha_x$ and $\alpha_z$, can be obtained by changing the potential barrier between the corresponding sites, as shown in Figure \ref{fig:bosons}. These bosonic terms can be controlled independently from the fermionic atoms as they are created by independent optical lattices (see Appendix \ref{AppE} for details). Pairing terms of the form $\alpha^\dagger_z \alpha^\dagger_x+\alpha_z\alpha_x$ can be activated by resonances to molecular bound states as it has been recently demonstrated with homonuclear $^{39}$K-$^{39}$K or heteronuclear $^{41}$K-$^{87}$Rb mixtures~\cite{PhysRevA.99.033612,PhysRevLett.125.195302}. Population interactions of the form $\alpha^\dagger_z \alpha_z \alpha^\dagger_x \alpha_x $ can be accurately controlled by employing Feshbach resonances, e.g. between Rb atoms, as has already been demonstrated experimentally ~\cite{PhysRevLett.81.5109,PhysRevLett.103.265302} (see Figure \ref{fig:bosons}). Such resonances are very versatile as they can generate positive, negative or zero interactions, or change the collision rate by several orders of magnitude just by selecting appropriate atomic states and tuning appropriately the external magnetic field. Hence, it is plausible that the quantum gravity model coupled to Dirac fermions given in \eqref{GQFThamiltonian} can be experimentally realised with a mixture of bosonic and fermionic ultra-cold atoms in optical lattices. 

\section{Quantum signatures of gravity}

The optical lattice system given by \eqref{OLH} and \eqref{eq:bos} in the low energy limit simulates Dirac fermions coupled to massive gravitons \eqref{GQFThamiltonian}. The presence of simulated quantum effects of gravity can be witnessed by directly measuring the bosonic field or by measuring the effect it has on the fermionic field. The pure gravity Hamiltonian \eqref{gravhamiltonian} describes simple tunnelling and pairing terms between bosonic modes 1 and 2. The signatures of these couplings can be directly measured in the quantum correlations $\langle d_1^\dagger d_2^\dagger\rangle$ and $\langle d_1^\dagger d_2\rangle$ of the corresponding Bose-Einstein condensates \cite{horvath2017above}. We can establish how faithfully the simulating Hamiltonian \eqref{eq:bos} reproduces the desired gravitational Hamiltonian \eqref{gravhamiltonian} by determining the dependence of the correlations on the ``gravitational'' coupling $G$.

We present now how to identify the presence of a fluctuating gravitational field from the behaviour of the fermionic quantum correlations. In the absence of gravity, i.e. $G=0$, the interaction term of the Dirac field disappears giving rise to free fermions. If the fermions are coupled to a fluctuating gravitation field then interactions between them emerge. Hence, we can identify the presence of fluctuating geometry by identifying if the Dirac fermions are free or interacting. The distinction between the two can happen by testing the applicability of Wick's theorem, i.e. testing the decomposition of four-point quantum correlations in terms of two-point correlations. Such two- and four-point correlation measurements are routinely realised in cold atom experiments. Hence, our scheme provides a direct quantum signature of gravity in terms of experimentally feasible components.

The partition function of the system at temperature $T$ is given by
\beq
Z=\int \mathcal D \, \xi \mathcal \, \mathcal D \pi \mathcal \, \mathcal D  \psi \mathcal D  \bar \psi \, \exp \left(-\frac{1}{k_B T}\int d^2 x\; \mathcal H\right),
\label{eq:part}
\eeq
where $\mathcal H= \psi^\dagger \; h({\boldsymbol p})\; \psi  + \mathcal H_\text{gr}$ and $k_B$ is the Boltzmann constant. To determine the behaviour of the fermions due to their interactions mediated by gravitational fluctuations we can integrate the bosonic part, i.e. $\xi$ and $\pi$, our from \eqref{eq:part} and derive the effective fermionic Hamiltonian. Integrating out the momenta, $\pi^i_a$, leads to an irrelevant global factor in $Z$. We can subsequently integrate out the bosonic field $\xi^a_i$, which up to an overall constant yields 
\beq
Z = \int \mathcal D  \psi \mathcal D  \bar \psi \exp\left(-\frac{1}{k_B T} \int d^2 x \;\mathcal H_\text{eff}\right),
\eeq
with the effective Hamiltonian
\beq
\mathcal H_\text{eff}=-i\bar \psi  \gamma^i \partial_i \psi  -\frac{4\pi G}{l^2 \mu^2 }\epsilon_{ab}\epsilon^{ij} {\cal J}_i^a {\cal J}_j^b ,
\label{eq:effHam}
\eeq
where ${\cal J}_i^a= \frac{i}{2l}\left(\bar \psi  \gamma^a \partial_i \psi- \partial_i\bar \psi  \gamma^a  \psi\right)$ is the fermionic current. As a result the partition function of the system is Hubbard-like, effectively describing interacting fermions with a coupling proportional to the gravitational constant $G$. The presence of such attractive fermion interactions can be measured in optical lattices by monitoring its Hubbard-like behaviour~\cite{landig2016quantum}. It is interesting to note that similarity of the interaction term in \eqref{eq:effHam} and the effective gravitational models considered in \cite{Diakonov:2011im,Volovik:2021wut}.

An alternative manifestation of interactions \eqref{eq:effHam} mediated by the gravitational field is on the fermionic correlations, which can be witnessed by testing the applicability of Wick's theorem. In the absence of gravitational fluctuations, i.e. for Newton's constant $G\to 0 $ in \eqref{expeandomega} or for fluctuations $\xi_i^a \to 0$, the effective partition function corresponds to free fermions, as seen by \eqref{eq:effHam}. For this case, Wick's theorem states that all four-point quantum correlators of the ground state can be exactly decomposed in terms of two-point correlators. Such four-point correlators of fermions can be expressed in terms of fermionic densities and two-point correlations and thus can be directly measured in cold atom experiments \cite{gross2017quantum}. When the interactions induced by the gravitational field are present, i.e. for $G\neq 0$, Wick's decomposition does not apply, leaving a difference that can be determined by measurements of fermionic correlations \cite{matos2021emergence}. In the perturbative regime considered here this difference gives a measure of the coupling $G$ between the Dirac fermions and the gravitational field.

\section{Conclusions}

Among the forces of nature gravity keeps its quantum aspects well hidden. This lack of experiential evidence hinders the theoretical understanding of quantum gravity and its unification with the rest of the fundamental forces within the Standard Model. Here we propose a way of simulating quantum signatures of massive gravity coupled to Dirac fermions in the laboratory. By building upon recent methods, developed for simulating scalar or gauge fields coupled to Dirac fermions, we are able to model Dirac fermions in the presence of fluctuating geometries, which is a unique characteristic of quantum gravity. This breakthrough was possible by encoding spacetime geometries intrinsically in the couplings of the system and then employing bosonic fields that fluctuate these couplings. The bosonic fields are appropriately designed in order to give rise to quantum a massive graviton akin to the magnetoroton emerging in  fractional quantum Hall liquids. As our model provides a direct link between microscopic system components and properties of the effective massive gravity, it sheds light into the mechanism behind the emergence of quantum gravitons in strongly interacting systems.

Beyond providing an example where massive gravitons emerge in a condensed matter system, our model can be simulated with optical lattices. The components used in our proposal, such as the fermionic honeycomb lattice or the bosonic condensates, can be realised in the laboratory with current ultra-cold atom technology. Quantum signatures of the emerging geometry can be witnessed in the effective fermionic interactions mediated by their coupling to fluctuating geometry, which can be directly probed in optical lattice experiments.

Our work opens up a host of various applications. It is intriguing to consider the behaviour of various quantum gravity and cosmology theories in $1+1$, $2+1$ or $3+1$ dimensions with or without fermions. Several open questions exist about the quantum aspects of gravity, cosmology and the physics of black holes that can be addressed within the framework presented here. We envision that our proposal will initiate a new line of investigations where interacting models that simulate gravity can be realised in the laboratory and guide theoretical investigations towards the understanding of quantum aspects of gravity in nature.

\section*{Acknowledgments} 

We would like to thank Almut Beige, Marc Henneaux, Matthew Horner and Paolo Maraner for inspiring conversations.
This project was partially funded by the EPSRC (Grant No.\,EP/R020612/1), by FNRS-Belgium (conventions
FRFC PDRT.1025.14 and IISN 4.4503.15), as well as by
funds from the Solvay Family. P.S-R. has received funding from the Norwegian Financial Mechanism 2014-2021 via the Narodowe Centrum Nauki (NCN) POLS grant 2020/37/K/ST3/03390. The data that support the findings of this study are available from the authors upon request.

\appendix

\section{Spin connection}
\label{AppA}
We consider a three-dimensional geometry whose metric tensor is defined by a set of dreibeins $e^A_\mu$
\beq
g_{\mu\nu} = \eta_{AB}e^A_\mu e^B_\nu.
\eeq
where here $A=(0,a=1,2)$ denotes Lorentz indices, $\mu=(t,i=x,y)$ stand for manifold indices, and $\eta={\rm diag}(-,+,+)$ is the Minkowski metric. Parallel transport is defined by the spin connection $\omega^A_\mu$, which can be obtained from the \emph{vielbein postulate}
\beq\label{app:vielbeinpost}
\partial_\mu e^A_\nu - \Gamma^\rho_{\mu\nu}e^A_\rho +\epsilon^{A}_{\;BC}\omega^B_\mu e^C_\nu=0,
\eeq
where $\Gamma^\rho_{\mu\nu}$ is the affine Christoffel connection and $\epsilon^{ABC}$ the Levi-Civita symbol ($\epsilon^{012}=1$). In this case, the gravitational fluctuations translate into fluctuations of the dreibein and the spin connection
\beq\label{app:expeandomega}
e^A_\mu = \bar e^A_\mu+ \frac{1}{8\pi G} \xi^A_\mu  \,,\qquad \omega^A_\mu = \bar \omega^A_\mu +\frac{1}{8\pi G}  v^A _\mu .
\eeq
We restrict the analysis to flat backgrounds with constant dreibein and vanishing spin connection. Therefore we set
\beq\label{app:bg}
\bar e^A_\mu = constant \,,\qquad \bar \omega^A_\mu =0.
\eeq
Furthermore, we consider torsionless geometries. In this case taking the antisymmetric part of \eqref{app:vielbeinpost} allows one to express $\omega^A_\mu$ in terms the dreibein and its derivatives. For the corresponding perturbations \eqref{app:expeandomega} one finds the relation
\beq
\epsilon^{\mu\nu\rho} \left( \partial_\nu \xi^A_{\rho} + \epsilon^A_{\;BC} \, \bar e^B_\nu   v^C_\rho \right) = 0,
\eeq
which can be used to find the form of the spin connection perturbations $v^A_\mu$
\beq\label{app:solv}
v^A_\mu = M^{AB}_{\mu\nu}
 \epsilon^{\nu\alpha\beta}\partial_\alpha \xi_{B\beta} \,,\quad M^{AB}_{\mu\nu}\equiv\frac{1}{\bar e} \left(\frac{1}{2} \bar e^A_\mu \bar e^B_\nu - \bar e^A_\nu \bar e^B_\mu\right), 
\eeq
where $\bar e = {\rm det} (\bar e^A_\mu)$. When restricted to metrics of the form \eqref{gaussian}, which means
\beq\label{app:gaugefixede}
\bar e^A_\mu= \begin{pmatrix}
1 & 0 \\
0 & \bar e^a_i 
\end{pmatrix}
\,,\qquad 
\xi^a_\mu= \begin{pmatrix}
0 & 0 \\
0 & \xi^a_i 
\end{pmatrix},
\eeq 
the components of the spin connection reduce to
\beq\label{app:gaugefixedw}
\bal
&v^0_t=-\frac{4\pi G}{\bar e }  \epsilon^{ij} \bar e^a_i \dot \xi_{aj} \,,\qquad v^0_i=-\frac{8\pi G}{\bar e \,} \epsilon^{jk} \bar e^a_i \partial_j \xi_{ak}  \,,\\
&v^a_t=0 \,,\qquad \qquad v^a_i= 8\pi G M^{ab}_{ij}
 \epsilon^{jk}\dot \xi_{bk}   \,.
\eal
\eeq

\section{Gravity action}
\label{AppB}

In order to describe gravitational fluctuations, we start with the Palatini action for $e^A_\mu$ and $\omega^A_\mu$
\beq
S_\text{gr} =\frac{1}{8\pi G}\int d^3 x \,\epsilon^{\mu\nu\rho} e^A_\mu \left(\partial_\nu \omega_{A\rho} + \frac{1}{2} \epsilon_{ABC}\; \omega^B_\nu \omega^C_\nu \right) .
\eeq
Using \eqref{app:expeandomega}, the action can be expanded in powers of the (2+1)-dimensional reduced Planck mass $1/8\pi G$ as
\beq\label{sgravexp}
S_\text{gr} [\xi]= \frac{1}{8\pi G} S_\text{gr} ^{(0)} + S_\text{gr} ^{(1)}  +8\pi G  \,S_\text{gr}^{(2)} .
\eeq 
Defining the background curvature and torsion,
\beq
\bal
&\bar R^A_{\mu\nu}= \partial_{[\mu}\bar\omega^A_{\nu]}+ \frac{1}{2} \epsilon^A_{BC}\bar\omega^B_{[\mu} \bar\omega^C_{\nu]}=0,\\
&\bar T^A_{\mu\nu} = \partial_{[\mu}\bar e^A_{\nu]} +  \epsilon^A_{BC} \bar\omega^B_{[\mu} \bar e^C_{\nu]}=0,
\eal
\eeq
the different terms in the action \eqref{sgravexp} can be written as
\beq\label{qaction}
\bal
S_\text{gr} ^{(0)} &= \frac{1}{2} \int d^3 x \,\epsilon^{\mu\nu\rho}   \bar e^A_\mu  \bar R_{A\nu\rho},  \\
S_\text{gr} ^{(1)} &=  \frac{1}{2}\int d^3 x \,\epsilon^{\mu\nu\rho}   \left( \xi^A_\mu  \bar R_{A\nu\rho},
+ \bar v^A_\mu  \bar T_{A\nu\rho}
 \right),
\\
S_\text{gr} ^{(2)} &=\int d^3 x \,\epsilon^{\mu\nu\rho}\left( \xi^A_\mu  \bar D_\nu v_{A\rho} +\frac{1}{2} \epsilon_{ABC} \;\bar e^A_\mu v^B_\nu  v^C_\rho \right).
\eal
\eeq
Considering a flat torsionless background implies that $\bar R^A_{\mu\nu}=0= \bar T^A_{\mu\nu}$. Replacing \eqref{app:solv} in \eqref{qaction} and using tensor notation then yields
\beq\label{fullgaction}
\bal
&S_\text{gr} [\xi]=- 4\pi G \int d^3 x M^{AB}_{\mu\nu}
\epsilon^{\mu\alpha\beta}\epsilon^{\nu\gamma\delta}  \partial_\alpha \xi_{A\beta}  \partial_\gamma \xi_{B\delta}.
\eal
\eeq
One can check that the action \eqref{fullgaction} boils down to the massless Fierz-Pauli action for $h_{\mu\nu}= \bar e_{A\mu} \xi^A_\nu + \bar e_{A\nu}\xi^A_\mu$. 

We are interesting in adding a mass $\mu$ to the geometry fluctuations $\xi^A_\mu$. We do so by means of the Fierz-Pauli mass term
\beq\label{app:FPmass}
4\pi G\mu^2\epsilon^{\mu\nu\rho}\epsilon_{ABC}\, \bar e^A_\mu \xi^B_\nu \xi^C_\rho \,.
\eeq
Thus, implementing \eqref{app:gaugefixede} and \eqref{app:gaugefixedw} in \eqref{fullgaction} and adding the term \eqref{app:FPmass} restricted to those conditions, we find the following effective massive gravitational action
\beq\label{gravactionfinalapp}
\bal
S_\text{gr}  [\xi]= -4\pi G \int&d^3 x \epsilon^{ij}\epsilon_{ab}  \bigg(   \dot \xi^a_{i} \dot \xi^b_{j} 
- \mu^2 \xi^a_i \xi^b_j\bigg).
\eal
\eeq
where we have defined $\epsilon^{ij}\equiv \epsilon^{0ij}$.

\section{Field theory Hamiltonian}
\label{AppC}
The Hamiltonian density associated to the Lagrangian in \eqref{fullaction} is obtained after a straightforward Legendre transformation.
\beq
\mathcal H= \Pi^\dagger \dot\psi +\dot \psi \Pi+ \pi_a^i \dot \xi^a_i- \mathcal L,
\eeq 
where the canonical momenta read
\beq
\bal
&\Pi^\dagger=\frac{\partial \mathcal L}{\partial \dot\psi}=i \psi^\dagger \,,\qquad
\Pi=\frac{\partial \mathcal L}{\partial \dot\psi^\dagger}=0\,,
\\
&\pi_a^{i}=\frac{\partial \mathcal L}{\partial \dot\xi^a_{i}}=-8\pi G \epsilon^{ij}\epsilon_{ab} \dot\xi^b_j ,
\eal
\eeq
and satisfy the Poisson brackets
\beq\label{app:poissonb}
\bal
&\left\{\psi_\alpha(x), \Pi^\dagger_\beta(y)\right\}=
\left\{\psi^\dagger_\alpha(x) , \Pi_\beta(y)\right\}=-\delta_{\alpha\beta}\delta^{(2)}(x-y),
\\
&\left\{\xi^a_i(x), \pi_b^j(y)\right\}=
\delta^a_b\delta^j_i\delta^{(2)}(x-y).
\eal
\eeq
The Hamiltonian reduces to
\beq
\label{app:QFThamiltonian}
H= \int  d^2 x\; \mathcal H=  \int  d^2 x\; \left[\psi^\dagger \; h(\boldsymbol p)\; \psi  + \mathcal H_\text{gr}\right] ,
\eeq
with the single particle hamiltonian $h(\boldsymbol p)$
\beq\label{app:singpart}
h(\boldsymbol p)=\frac{\gamma^0}{l} \left( \delta^i_a\gamma^a -\frac{8\pi G}{l } \xi ^i_a \gamma^a \right)(-i \partial_i ) +\frac{4i\pi G}{l^2}
 \partial_i \xi^i_a  \gamma^0  \gamma^a ,
\eeq
and the gravitational Hamiltonian $\mathcal{H}_\text{gr}$ given by
\beq\label{app:gravhamiltonian}
 \mathcal{H}_\text{gr} =-  \frac{1}{16\pi G}\epsilon_{ij} \epsilon^{ab} \pi_a^i  \pi_b^j-{4\pi G\mu^2}\epsilon^{ij} \epsilon_{ab}  \xi^a_i  \xi^b_j.
\eeq

Since our analysis considers geometry fluctuations $\xi^a_i$ that are diagonal, we define quantum operators of the form
\beq
\xi^a_i = \begin{pmatrix} 
 \xi^1_x & 0\\
 0 & \xi^2_y 
\end{pmatrix}
\,,\qquad
\bal
&\xi^1_x= \frac{1}{\sqrt 2}(q_1^\dagger +q_1)\,,\\&\xi^2_y= \frac{1}{\sqrt 2}(q_2^\dagger +q_2),
\eal
\eeq
where the operators $q_a$ and $q^\dagger_a$ $(a=1,2)$ satisfy the commutation relations
\beq\label{commq}
\bal
&[q_a(x), q_b^\dagger(y) ]=\delta_{ab}\delta^{(2)}(x-y),
\\
&[ q_a(x), q_b(y) ]=0=[q_a^\dagger(x), q_b^\dagger(y) ].
\eal
\eeq
Thus, the canonical Poisson brackets \eqref{app:poissonb} are promoted commutators $\{\;,\;\}\rightarrow -i[\;\,\;]$, which then fixes the form of the momenta $\xi_a^i$ to be
\beq
\pi_a^i = \begin{pmatrix} 
 \pi_1^x & 0\\
 0 & \pi_2^y 
\end{pmatrix}
\,,\qquad
\bal
&\pi_1^x= -\frac{i}{\sqrt 2 }(q_1^\dagger -q_1)\,,\\&\pi_2^y=-\frac{i}{\sqrt 2 }(q_2^\dagger -q_2).
\eal
\eeq
The flat background spatial metric is taken as $g_{ij}=l^2\delta_{ij}$ with $l$ an arbitrary constant, which implies
\beq
\bar e^a_i = l \delta^a_i
\,,\qquad
\bar e^i_a = \frac{1}{l} \delta^i_a \,,
\eeq
and thus the gamma matrices on this background geometry reduce to
\beq
\gamma^t = \gamma^0 
\,,\quad
\gamma^i = \frac{1}{ l }  \delta^i_a \gamma^a.
\eeq
From this expressions we see that the single particle Hamitlonian \eqref{app:singpart} can be written as
\beq\label{HspOp}
\bal
 h(\boldsymbol p)&= 
   \left( \frac{1}{ l }- \frac{ 4\sqrt2 \pi G }{  l ^2}  (q_1^\dagger +q_1) \right)\gamma^0\gamma^1(-i\partial_x )  
\\&
+\left( \frac{1}{ l } - \frac{ 4\sqrt2 \pi G}{  l ^2} (q_2^\dagger +q_2)\right)\gamma^0\gamma^2 (-i\partial_y)    \\
& +\frac{2\sqrt2 i\pi G}{  l ^2}\left(
\partial_x (q_1^\dagger+q_1) \gamma^0\gamma^1
+
\partial_y (q_2^\dagger+q_2) \gamma^0  \gamma^2 \right),
\eal
\eeq
whereas the gravitational Hamiltonian \eqref{app:gravhamiltonian}  takes the form
\beq\label{HgrOp}
\mathcal H_{\rm gr} =\frac{1}{16\pi G } (q_1^\dagger -q_1)(q_2^\dagger -q_2)-4\pi G \mu^2 (q_1^\dagger +q_1)(q_2^\dagger +q_2).
\eeq
In the following, we show how to translate \eqref{app:QFThamiltonian} into an optical lattice Hamiltonian.

\section{ Optical Lattice Hamiltonian}
\label{AppD}

We start considering an optical lattice with unequal tunnelling couplings
\beq
H_{\rm latt}=\sum_{ \boldsymbol i} \left( 
J_x a_{\boldsymbol i}^\dagger b_{\boldsymbol i+\boldsymbol  n_1}
+J_y a_{\boldsymbol i}^\dagger b_{\boldsymbol i+\boldsymbol n_2}
+J_z a_{\boldsymbol i}^\dagger b_{\boldsymbol i}
\right)
+ {\rm h.c.},
\eeq
where $\boldsymbol{i}=(i_x,i_y)$ denotes the position of the unit cells (see Figure \ref{fig:lattice}).
Expanding the operators in Fourier modes and defining
\beq
\psi_{\boldsymbol k} = \begin{pmatrix} 
 a_{\boldsymbol k}\\
b_{\boldsymbol k}
\end{pmatrix},\quad
\tilde h({\boldsymbol k} )=\begin{pmatrix} 
 0 & f( \boldsymbol k)\\
 f^*( \boldsymbol k) &0 
\end{pmatrix},
\eeq
where $f( \boldsymbol k)= J_x e^{-i \boldsymbol k \cdot  \boldsymbol n_1}+ J_y e^{-i \boldsymbol k \cdot  \boldsymbol n_2}+J_z$, we find
\beq
H_{\rm latt}= \sum_{ \boldsymbol k} \psi^\dagger_{\boldsymbol k} \tilde h( \boldsymbol k ) \psi_{ \boldsymbol k }.
\eeq
We consider the special case $J_x=J_y$. The Fermi points, $\boldsymbol P_{\pm}$, are defined by
\beq
f( \boldsymbol P_{\pm} ) =0 \Rightarrow  \boldsymbol P_{\pm}= \pm
\begin{pmatrix} 
\frac{2}{\sqrt3}\arccos (-J_z/J_x)\\
0 
\end{pmatrix}.
\eeq
Now, we expand $f( \boldsymbol k ) $ around the Fermi points
\beq
f( \boldsymbol P_\pm+ \boldsymbol p  )= \boldsymbol p \cdot \nabla f( \boldsymbol P_\pm)= A_{\pm}  p_x + B_{\pm}   p_y ,
\eeq
where we have defined
\beq
A_{\pm}= \mp \frac{\sqrt3}{2} \sqrt{4J_x^2-J_z^2},\quad
B_{\pm} = -\frac{3}{2} J_z .
\eeq
Expanding the Hamiltonian $H_{\rm latt}$ around the Fermi point yields
\beq
\bal
H_{\rm latt}&= \sum_{ \boldsymbol p} \begin{pmatrix} 
a^\dagger_+\;\; b^\dagger_+
\end{pmatrix}
(A_+ \sigma^1  p_x-B_+ \sigma^2  p_y) \begin{pmatrix} 
a_+\\
b_+
\end{pmatrix}
\\&+
\sum_{ \boldsymbol p}
\begin{pmatrix} 
a^\dagger_- \;\; b^\dagger_-
\end{pmatrix}
(A_- \sigma^1  p_x-B_- \sigma^2  p_y)
\begin{pmatrix} 
a_-\\
b_-
\end{pmatrix}
.
\eal
\eeq
Defining the four spinor and the gamma matrices
\beq
\psi_{\boldsymbol p} =
\begin{pmatrix} 
a_+\\
b_+\\
b_-\\
a_-
\end{pmatrix}
,\quad
\gamma^0=\begin{pmatrix} 
 0 &-\boldsymbol 1\\
 \boldsymbol1 &0 
\end{pmatrix},\quad
\gamma^i=\begin{pmatrix} 
 0 &\sigma^i \\
\sigma^i &0 
\end{pmatrix},
\eeq
with $\sigma^i=(\sigma^1,\sigma^2)$ Pauli matrices. Then we find $H_\text{latt} = \int  d^2 x\; 
 \psi^\dagger \;\tilde h({\boldsymbol p}) \;\psi$ with the single particle Hamiltonian
\beq
\tilde h(\boldsymbol p)=\frac{\sqrt3}{2} \sqrt{4J_x^2-J_z^2}\,\gamma^0\gamma^1p_x  
+ \frac{3}{2}J_z \gamma^0\gamma^2p_y .
\eeq
where, for convenience, we have flipped the orientation of the $y$ axis, i.e. $p_y\shortrightarrow -p_y$. In order to ensure hermiticity of the full Hamiltonian, the momentum operator is defined as $p_i=-i(\overrightarrow\partial_i-\overleftarrow\partial_i)/2$. Thus we find
\beq\label{app:h}
\bal
\tilde h(\boldsymbol p)&=  \frac{\sqrt3}{2} \sqrt{4J_x^2-J_z^2}\,\gamma^0\gamma^1(-i\partial_x)  
+ \frac{3}{2}J_z \gamma^0\gamma^2(-i\partial_y) \\
&+ \frac{\sqrt3i}{2} \partial_x \left(\sqrt{4J_x^2-J_z^2}\,\right)\gamma^0\gamma^1+ \frac{3i}{2}\partial_y J_z \gamma^0\gamma^2 .
\eal
\eeq
We can now generalise this Hamiltonian by considering the couplings $J_x$ and $J_z$ as position dependent, varying slowly compared to the lattice spacing. Next, we add a bosonic self-interacting Hamiltonian $\mathcal H_\text{boson}$,
\beq\label{app:boson}
\bal
\mathcal H_\text{boson} &= \frac{ 1}{24\pi G}(\alpha_z^\dagger -\alpha_z)\left(\sqrt2 (\alpha_x^\dagger -\alpha_x)-\frac{1}{2}
(\alpha_z^\dagger -\alpha_z)\right)\\
&+\frac{8\pi G \mu^2}{3}\left(\alpha^\dagger_z \alpha_z+ 
\alpha^\dagger_x \alpha_x\right)\\
& 
-\frac{ 256\pi^3G^3\mu^2}{3 l^2} \alpha^\dagger_z \alpha_z \left(\alpha^\dagger_x \alpha_x 
-\frac{1}{2}\alpha^\dagger_z \alpha_z\right),
\eal
\eeq
where $\alpha_x$ and $\alpha_z$ are bosonic modes at the edges of the lattice that control the tunneling of fermions in the $x$ and in the $z$ direction, respectively (see Figure \ref{fig:superlattices}). The optical lattice Hamiltonian that we will consider is then
\beq\label{app:OLH}
\bal
H_{\rm latt}= \int  d^2 x\; &\bigg[
 \psi^\dagger \;\tilde h(\boldsymbol p)\; \psi    +\mathcal H_\text{boson} 
 \bigg].
\eal
\eeq
In order to show that \eqref{app:OLH} can be mapped to \eqref{app:QFThamiltonian}, first we consider the following form of the couplings
\beq
J_m= \Delta_m D_m \left(D_m +d_m^\dagger +d_m \right)\,, \quad m=x,z .
\eeq
where the operators $d_m$ satisfy the commutation relations
\beq
\bal
&[d_m(x), d_n^\dagger(y) ]=\delta_{mn}\delta^{(2)}(x-y),
\\
&[ d_m(x), d_n(y) ]=0=[d_m^\dagger(x), d_n^\dagger(y) ],
\eal
\eeq
Note that this is a continuum version of the commutation relations given below Eq. \eqref{defalpham}. By considering
\beq\label{app:DDeltas}
\bal
&D_{x}=-{l\over 4 \pi G}\,,\qquad \Delta_x=\frac{32\pi^2G^2}{3 l^3 },
\\&
D_{z}=-{\sqrt{2}l \over 8\pi G} \,,\qquad\Delta_{z}=\frac{64\pi^2G^2}{3 l^{3}},
\eal
\eeq
one finds that the different terms in $\tilde h(\boldsymbol p)$ given in \eqref{app:h} can be written in the form
\beq
\bal
\frac{3}{2} J_z &= \frac{1}{ l } - \frac{ 4 \sqrt2 \pi G}{  l ^2}\left(d_z^\dagger +d_z\right),
\\
\frac{\sqrt{3}}{2}\sqrt{4J_{x}^{2}-J_{z}^{2}}
&\approx \frac{1}{ l } -\frac{16 \pi G}{ 3 l ^2 } \left(d_x^\dagger + d_x\right)
\\&
\hskip.8truecm+\frac{4 \sqrt 2 \pi G }{3  l ^2 }\left(d_z^\dagger +d_z \right),
\eal
\eeq
where in the last equality we have used the approximation
\beq\label{appdD}
\left| \left\langle d^\dagger_m + d_m \right\rangle\right| \ll \left|D_m\right|,
\eeq
valid for bosonic states in the weak fluctuation regime.
Finally, we consider the operator redefinition
\beq
\label{app:redefop}
q_1= \frac{2\sqrt2}{3} d_x -\frac{1}{3} d_z \,,\qquad q_2=d_z,
\eeq
which satisfy the canonical commutation relations given in \eqref{commq}. This transformation maps the single particle Hamiltonian $\tilde h(\boldsymbol p)$, given \eqref{app:h}, into the corresponding Hamiltonian $h(\boldsymbol p)$ describing fermions coupled to dreibein fluctuations given in \eqref{HspOp}.

Now we turn our attention to the bosonic Hamiltonian \eqref{app:boson}, where the operators $\alpha_m$ are defined as
\beq
\alpha_m = D_m+ d_m.
\eeq
In this case we reverse the procedure and show that, up to an irrelevant additive constant, the gravitational Hamiltonian $\mathcal H_{\rm gr}$ given in \eqref{HgrOp} can be mapped to \eqref{app:boson}.  This can be directly shown by noticing that
\beq
\bal
 (d^\dagger_m -d_m ) (d^\dagger_n -d_n )
&= (\alpha^\dagger_m -\alpha_m)  (\alpha^\dagger_n -\alpha_n ),\\
(d^\dagger_m +d_m ) (d^\dagger_m +d_m )
&\approx 
\frac{\alpha^\dagger_m \alpha_m \alpha^\dagger_n \alpha_n }{D_m D_n}-\frac{D_m}{D_m}\alpha^\dagger_n \alpha_n\\
&-\frac{D_n}{D_m}\alpha^\dagger_m \alpha_m +D_m D_n,
\eal
\eeq
and using \eqref{app:DDeltas} and \eqref{app:redefop}. Note that in the last relation we have used $\alpha^\dagger_m \alpha_m \approx D_m^2 + D_m (d_m^\dagger +d_m)$, which holds in the approximation \eqref{appdD}.  

\section{Optical lattice realisation of the gravitational Hamiltonan}
\label{AppE}

In order to realise the bosonic Hamiltonian \eqref{eq:bos}, we consider the Hamiltonian for bosonic atoms, described by a bosonic mode $\Phi({\bf x})$, in an optical lattice potential $V_0$ and a slowly varying external trapping potential $V_T$:
\beq
\bal
H=&\int d^3x \,\, \Phi^\dagger ({\bf x}) \left[-\frac{\nabla ^2}{2m} +V_0
({\bf x}) + V_T ({\bf x}) \right] \Phi({\bf x})
\\
& + \frac{2 \pi a_s  ^2}{m} \int d^3 x \,\, \Phi^\dagger ({\bf x})
\Phi^\dagger ({\bf x}) \Phi({\bf x}) \Phi({\bf x})
\eal
\eeq
where $a_s$ is the $s$-wave scattering length and $m$ is the mass of the atoms \cite{PhysRevLett.81.3108}. 
For single atoms the energy eigenstates are Bloch wave functions and an appropriate superposition of Bloch states yields a set of Wannier functions which are well localized on the individual lattice sites. Expanding the field operator in the Wannier basis $w({\bf x})$ in the following way,
\beq
\Phi({\bf x}) =\sum_i \alpha_i \phi_i ({\bf x}), \quad \phi_i ({\bf x})=w({\bf x}-{\bf x}_i),\quad [\alpha_i, \alpha_j^\dagger ]=\delta_{ij}
\eeq
one finds the Bose-Hubbard Hamiltonian
\beq
H=-\sum_{<ij>} t_{ij}\; \alpha_i^\dagger \alpha_j + \sum_{<ijkl>} U_{ijkl}\; \alpha_i^\dagger \alpha_j^\dagger \alpha_k \alpha_l,
\label{ham1}
\eeq
where $<\!\!\cdots\!\!>$ denotes summation only over nearest neighbours. In \eqref{ham1} we have defined the overlap integrals
\beq
\bal
t_{ij}&=\int d^3 x\, \phi^*_i({\bf x}) \left(-\frac{\nabla^2}{2m} + V_0({\bf x})+ V_T({\bf x})\right)\phi_j({\bf x})
\\
U_{ijkl}&=\frac{4 \pi a_s  ^2}{m} \int d^3 x \,\phi^*_i({\bf x})\, \phi^*_j({\bf x}) \,\phi_k({\bf x})\,\phi_l({\bf x}),
\eal
\eeq
where $t_{ij}$ represents the tunnelling couplings (or the hopping matrix elements), whereas $U_{ijkl}$ stands for the strength of the on-site repulsion of two atoms.
We arrange the modes to be defined on a diamond lattice with indices $i=(\boldsymbol i,m)$, $j=(\boldsymbol j,n)$, etc, as shown in Figure \ref{fig:lattice}. For particular values of $t_{ij}$ and $U_{ijkl}$, e.g. by changing the laser intensity or employing Feshbach resonances that control the scattering length, it is possible to tune appropriately the couplings and reproduce the gravitational Hamiltonian \eqref{eq:bos}.


\end{document}